\documentclass{appolb}
\usepackage{epsfig}

\begin{document}
\eqsec  
\title{Why and how to use a differential equation method to calculate 
 multi-loop integrals %
\thanks{Presented by H. Czy{\.z} 
 at XXV International Conference on
  Theoretical Physics ``Particle Physics and Astrophysics 
 in the Standard Models
  and Beyond'', Ustro{\'n}, Poland, September 2001}
}

\author{M. Czachor 
\address{Institute of Physics, University of Silesia,
PL-40007 Katowice, Poland.}
\and
 H. Czy{\.z}
\address{Institute of Physics, University of Silesia,
PL-40007 Katowice, Poland and Institute of Advanced Study, 
University of Bologna, I-40138 Bologna}
}
\maketitle
\begin{abstract}
 A short pedagogical introduction to a differential method used to
 calculate multi-loop scalar integrals is presented. As an example 
 it is shown how to obtain, using the method, large mass expansion of the 
 two loop sunrise master integrals. 
\end{abstract}
  
\section{Introduction}
\newcommand{\labbel}[1]{\label{#1}} 
\newcommand{\dnk}[1]{ \frac{d^nk_{#1}}{(2\pi)^{n-2}} } 
\newcommand{\verso}[1]{ {\; \buildrel {n \to #1} \over{\longrightarrow}}\; } 
\newcommand{\F}[1]{F_{#1}(n,m_1^2,m_2^2,m_3^2,p^2)} 
\newcommand{\Ref}[1]{(\ref{#1})}

 Precision measurements have become one of the central issues
 in present particle physics allowing to test Standard Model
 and its extensions with unprecedented accuracy.
 To confront them with theoretical predictions it is necessary to know
 two (or more) loop radiative corrections to the measured physical
 quantities. Despite enormous effort of the theoretical physics
 community in this field and existence of many partial results a universal 
 method beyond first loop was not developed till now. While the
 planned linear colliders running in gigaZ mode will push the experimental
 accuracy even further.    

One of the promising new directions in the field of multi-loop
calculation is the  differential equations method.
A differential equations method, based on mass derivatives, 
has been proposed in \cite{Kotikov}. In that 
approach, amplitudes with a single non vanishing mass \( m \) are 
expressed as a suitable integral of the corresponding massless amplitudes, 
which are taken as known. More systematic studies making use not only
 of masses, but also Lorentz invariants, as independent variables
 were initiated in \cite{Ett} and then continued in
 \cite{CCLR1}-\cite{MR}. In \cite{CCLR1} small and large \(p^2\) expansions
 were obtained for sunrise type master integrals. In \cite{CCLR2} the same
 type of expansions was obtained for the two loop two point four denominator
 master integral.
 Analytical results for 
 pseudothreshold \cite{CCR1} and threshold \cite{CCR2} expansions 
 of the sunrise master integrals were obtained subsequently. 
 Another application of the method is the calculation of the massless off shell
 double box contributing to \(\gamma^* \rightarrow\) 3-jet process.
  The differential equations were presented
 in \cite{GR1} and subsequently 
 master integrals calculated for planar \cite{GR3} and
 non-planar \cite{GR4} topologies.
 In \cite{MR} and \cite{B} efforts towards getting 4-loop corrections to
  \(g-2\) and 2-loop corrections to
 Bhabha scattering correspondingly were started. 
 The power of the method, besides being relatively simple and
 based mostly on algebraic manipulations, is that
 its mathematical basis was developed long ago. As it will be shown
 in section 3 the master integrals ever satisfy a system of linear
 differential equations. Theory of such systems of differential equations
 was developed in XIX and at the beginning of XX century, so now it is
 a textbook knowledge (see for example \cite{I}). All that helps a lot,
 making possible mathematical rigor without big effort.

 This paper is organized as follows. In the next section some preliminary 
 definitions are given and integration by part identities \cite{ChetTka}
  shortly presented.
 In section 3 it is shown how to get differential equations once
 the master integrals are identified. In section 4 it is shown how
 to obtain large mass expansion of the sunrise master integrals. It is
 the first application of the differential equation method to the calculation
 of the mass expansions. In section 5 a short summary is presented. All
 algebraic manipulations were performed using FORM \cite{FORM}.

\section{Preliminaries}

 For a presentation how the differential equation method works in practice
 in a nontrivial case,
 but in the same time still not requiring large algebraic
 manipulations, we have chosen the two point two loop sunrise graph.
 A family of scalar integrals associated with that graph is defined by

\begin{eqnarray} 
  A(n,m_1^2,m_2^2,m_3^2,p^2,-\alpha_1,-\alpha_2,-\alpha_3,\beta_1,\beta_2) 
      &=& \nonumber \\  
      && {\kern-220pt} \int \dnk{1} \int \dnk{2} \; 
      \frac{ (p\cdot k_1)^{\beta_1} (p\cdot k_2)^{\beta_2} } 
           { (k_1^2+m_1^2)^{\alpha_1} (k_2^2+m_2^2)^{\alpha_2} 
             ( (p-k_1-k_2)^2+m^2_3 )^{\alpha_3} } \ \ ,
 \nonumber \\
\labbel{1} \end{eqnarray} 

\noindent
where \(m_i\) (\(i=1,2,3\) ) are the masses associated with internal lines,
\(p\) is the external momentum, \(k_1,k_2\) are loop momenta and 
 \(\alpha_i,\ \ (i=1,2,3)\), \(\beta_j,\ \ (j=1,2)\) are integer numbers.
 The integrals are to be performed in \(n\)-dimensional Euclidean space.
 It implies we have used dimensional regularization 
 and have performed Wick rotation.
 The scale parameter \(\mu\) associated with dimensional regularization
 has been set to 1. Final results can be easily rewritten in Minkowski space
 by changing \(p^2\rightarrow-p^2\).  

 Not all of the integrals from the class \Ref{1} are independent.
 By means of a very simple but powerful method of integration by parts
 identities \cite{ChetTka} one can find relations between them. 
 In this particular case one uses the relations

\begin{eqnarray} 
 \kern-8pt     \int d^nk_i
 \frac{\partial}{\partial (k_{i})_{\mu}}\Biggl[  
      \frac{v_{\mu} \ \  (p\cdot k_1)^{\beta_1} (p\cdot k_2)^{\beta_2} } 
           { (k_1^2+m_1^2)^{\alpha_1} (k_2^2+m_2^2)^{\alpha_2} 
             ( (p-k_1-k_2)^2+m_2^3 )^{\alpha_3} } \Biggr] = 0 \ \ ,\nonumber \\
\labbel{1a} \end{eqnarray} 

\noindent
where \(i=1,2\), while \(v_{\mu}\) denotes one of the momenta \(p,k_1\) or
 \(k_2\). It is crucial that in this way one gets a system of linear equations
 satisfied by the integrals \Ref{1} with a non-homogeneous terms given
 by integrals with lower number (in this case one) of denominators.
 This general property requires that one should start to solve a given
 problem from calculation of the integrals with smallest possible number
 of denominators ( or if one is lucky enough, one 
 can find them in the literature). The integrals to be calculated first for
 the sunrise problem can be expressed just by one integral

\begin{eqnarray} 
 T\left(n,m^2\right)=
 \int \frac{d^n k}{(2 \pi)^{n-2}} \frac{1}{k^2+m^2}
 = {\frac{m^{n-2}}{(n-2)(n-4)}} C(n) \ \ ,
 \labbel{a1}
 \end{eqnarray} 

\noindent
where

\begin{eqnarray} 
 C(n)= \left(2 \sqrt{\pi} \right)^{(4-n)} \Gamma\left(3-\frac{n}{2}\right) 
 \ \ \ \  {\mbox {\rm and}} \ \ \ \   C(4)=1 \ \ .
\labbel{a11}
 \end{eqnarray} 

\noindent
With help of \Ref{1a} one finds 
 that only four independent
integrals exist within the family \Ref{1}\cite{Tarasov}. We choose them as

\begin{eqnarray} 
  \F{0} &=& A(n,m_1^2,m_2^2,m_3^2,p^2,-1,-1,-1,0,0) \ , \nonumber \\ 
  \F{1} &=& A(n,m_1^2,m_2^2,m_3^2,p^2,-2,-1,-1,0,0) \ , \nonumber \\ 
  \F{2} &=& A(n,m_1^2,m_2^2,m_3^2,p^2,-1,-2,-1,0,0) \ , \nonumber \\ 
  \F{3} &=& A(n,m_1^2,m_2^2,m_3^2,p^2,-1,-1,-2,0,0) \ . 
\labbel{2} \end{eqnarray} 

Thereafter we will not write explicitly the arguments of the functions
 \(\F{i} \equiv F_i \). The independence of the function \(F_i\) means
 that none of them can be expressed by a linear combination of the
 others and polynomials of the
  function \(T\) \Ref{a1} with coefficients in the
 form of a ratio of polynomials in \(p^2,m_1^2,m_2^2,m_3^2\). However
 an obvious relation occurs

 \begin{eqnarray} 
 F_i = -\frac{\partial}{\partial m_i^2} F_0 \ \ \ i=1,2,3.
\labbel{2a} \end{eqnarray} 

\section{How to get differential equations for master integrals}

Having a limited number of integrals to deal with, which we will call
master integrals, we can write 
differential equations they obey. It is as simple as to differentiate
 the given integral and then by means of the integration by part
 identities express the result by the master integrals and the known
 function \(T\). To illustrate as it works we write

\begin{eqnarray} 
  &&p^2 {\frac{\partial} {\partial p^2}}F_0 = \nonumber \\
 &&{\frac{1}{2}}  p_\mu {\frac{\partial} {\partial p_\mu}}
 \int {\frac{d^n k_1 d^n k_2}{(2\pi)^{2n-4}}} 
 {\frac{1}{ (k_1^2+m_1^2) (k_2^2+m_2^2)
             ( (p-k_1-k_2)^2+m_3^2 ) }} = \nonumber \\
&&\int {{d^n k_1 d^n k_2}\over{(2\pi)^{2n-4}}} 
 {\frac{-p^2+p\cdot k_1+p\cdot k_2}
  { (k_1^2+m_1^2) (k_2^2+m_2^2)
             ( (p-k_1-k_2)^2+m_3^2 )^2 }} \ \ .
 \labbel{a4}
 \end{eqnarray} 

The last expression is nothing but a linear combination of three integrals
from the family \Ref{1}, so one can express each of them as 
a linear combination of the master integrals. This gives after
a short algebraic calculation \cite{CCLR1}

\begin{eqnarray} 
p^2 {\frac{\partial} {\partial p^2}} F_0 =
(n-3)F_0 + m_1^2F_1+m_2^2F_2+m_3^2F_3 \ \ . 
 \labbel{a5}
 \end{eqnarray} 

Similarly one finds \cite{CCLR1}

\begin{eqnarray} 
 p^2 \ D(m_1^2,m_2^2,m_3^2,p^2) 
  \ {\frac{\partial}{\partial p^2}} F_i =
 \sum_{j=0}^{3} M_{i,j}  F_j + T_i  \ \ , \ i=1,2,3
 \labbel{a6}
 \end{eqnarray} 

\noindent
where explicit form of functions \(T_i\) (which can be expressed by
the function \(T\) \Ref{a1}) and \(M_{i,j}\) (polynomials of 
\(p^2, m_1^2,m_2^2,m_3^2\) ) can be found in \cite{CCLR1}.
 The function \(D\) is defined by

 \begin{eqnarray} 
D(m_1^2,m_2^2,m_3^2,p^2)&=&
  \left(p^2+(m_1+m_2+m_3)^2\right) \ 
                        \left(p^2+(m_1+m_2-m_3)^2\right) \nonumber \\ 
     &&\kern-30pt  \cdot \left(p^2+(m_1-m_2+m_3)^2\right) \ 
                        \left(p^2+(m_1-m_2-m_3)^2\right) \ ,
  \labbel{a7}
 \end{eqnarray} 

\noindent
and is equal to zero at the three pseudothresholds and at the threshold of
 the master integrals.

 Differential equations with \(m_i\) as an independent variables
 can be found in a similar way. They are presented in the Appendix
 of \cite{CCLR1} or can be obtained from the formulae presented there
 by appropriate permutations of the masses \(m_i, \ \ i=1,2,3\).

The property that the derivative of a given master integral is a linear
 combination of the master integrals plus terms with smaller number of
 denominators is obviously a general property of all possible multi-loop scalar
 integrals. It is valid due to linearity
 of the integration by parts identities and the differentiation operation itself,
 and also due to the form of the integrands,
  which are ever ratios of polynomials
 of masses and Lorentz invariants.

\section{An application:
 Large mass expansion of the master integrals}

 Let us assume that one of the square of the masses, say \(m_3^2\), 
 is much larger then \(m_1^2,m_2^2\) and \(|p^2|\). The general
 form of the expansion in that region can be written \cite{I} as 

\begin{eqnarray} 
F_0 =\sum\limits_{\alpha \in A }(m_3^2)^{\alpha}
\sum\limits_{k=0}^{\infty}F_k^{(\alpha)}
{\frac{1}{{(m_3^2)^{k}}}} \ \ ,
\labbel{m1} \end{eqnarray} 

\noindent
where A is a finite set of numbers, whose differences are not equal
 to an integer number. Other master integrals are related to \(F_0\)
 by \Ref{2a}.
 The allowed values of \(\alpha\) can be found
 from the system of equations (with \(m_3^2\) as an independent variable)
 itself. One substitutes the \(F_i\) in the system with its expansions
 and by examining the coefficients of the highest powers in \(m_3^2\)
 (they have to be equal to zero) one finds allowed values of \(\alpha\)'s.
 In this particular case

\begin{eqnarray} 
 A=\Biggl\{-1, \ \ {\frac{n-4}{2}},\ \ n-3, \ \ {\frac{3}{2}}n-4\Biggr\} 
 \equiv \Biggl\{r, \ \ s_1, \ \ s_2, \ \ s_3 \Biggr\} \ \ ,
\labbel{m2} \end{eqnarray} 

\noindent
where a shorthand notation was introduced for different values of \(\alpha\)'s.
The series with an integer power of \(\alpha\) is called the regular series,
 while the other are called singular series. The singular parts are
 sources of logarithmic terms when expanded around \(n=4\), the value
 of \(n\)
 we are interested in.
 
 Not always all parts of the expansion corresponding to 
 allowed values of \(\alpha\)'s are actually present in the solution.
 That depends on the initial conditions and the regularity 
 of the \(F_i\) at \(p^2=0\) is crucial for their properties \cite{CCLR1}.
 We will see that also in the presented below example. It reflects the fact
 that the differential equations can be satisfied by a wider class of
 functions, not only by the master integrals. It means also that
 usually one has to calculate the master integrals for special values
 of the parameters by other means then the differential equations
 to fix constants of integration. That however is always simpler
 then the general case.

 Having the allowed values of \(\alpha\) one can try to calculate
 coefficients in the expansion \Ref{m1}.
  The crucial point is to find the first coefficient in
 each of the series as the others can be found by solving a system
 of linear algebraic equations (in this case system of 3 linear equations).
 Two of the coefficients are fixed by non-homogeneous terms
 in the differential equations and read

\begin{eqnarray} 
F^{r}_0={\frac{C^2(n)}{(n-4)^2(n-2)^2}}(m_1^2m_2^2)^{\frac{n-2}{2}}
 \ \ ,
\labbel{m3} \end{eqnarray} 

\begin{eqnarray} 
F^{s_1}_0=
 {-\frac{C^2(n)}{((n-4)(n-2))^2}}
 \biggr((m_1^2)^{\frac{n-2}{2}}+
(m_2^2)^{\frac{n-2}{2}}\biggr)
\labbel{m4} \end{eqnarray} 

The other two cannot be fixed this way as in the non-homogeneous part of the
 equations there is no term \(\sim (m_3^2)^{s_2}\) or \(\sim (m_3^2)^{s_3}\).
 One can find however the following  relations between the next to leading 
 and the leading terms in the expansion

\begin{eqnarray} 
F_1^{s_2}&=&-(n-3)\biggr[(m_1^2+m_2^2+p^2)-{\frac{4}{n}}p^2\biggr]F_0^{s_2}
 \nonumber \\
&&\kern-17pt
+ {\frac{4}{n}}(n-3)m_1^2(m_1^2+p^2){\frac{\partial}{\partial m_1^2}}F_0^{s_2}
+ {\frac{4}{n}}(n-3)m_2^2(m_2^2+p^2){\frac{\partial}{\partial m_2^2}}F_0^{s_2}
 \nonumber \\
\frac{\partial}{\partial m_1^2}F_1^{s_2}&=&
(n-3)\biggr[(p^2-m_1^2+m_2^2)\frac{\partial}
 {\partial m_1^2}F_0^{s_2}-F_0^{s_2}\biggr]
 \nonumber \\
\frac{\partial}{\partial m_2^2}F_1^{s_2}&=&
(n-3)\biggr[(p^2+m_1^2-m_2^2)\frac{\partial}
 {\partial m_2^2}F_0^{s_2}-F_0^{s_2}\biggr]
\labbel{p11}\end{eqnarray}

Using differential equations \Ref{a5} and \Ref{a6}, the expression \Ref{m1}
and the above relations one finds that \(F_0^{s_2}\) satisfies
 the following system of differential equations

\begin{eqnarray}
-m_1^2\frac{\partial}{\partial m_1^2}F_0^{s_2}
-m_2^2\frac{\partial}{\partial m_2^2}F_0^{s_2}
-p^2\frac{\partial}{\partial p^2}F_0^{s_2}=0
 \   \  ,
\labbel{p12}\end{eqnarray}

\begin{eqnarray}
\frac{n-2}{2}\frac{\partial}{\partial m_1^2}F_0^{s_2}
+p^2\frac{\partial}{\partial m_1^2}\frac{\partial}{\partial p^2}F_0^{s_2}=0
 \   \  , \nonumber \\
\frac{n-2}{2}\frac{\partial}{\partial m_2^2}F_0^{s_2}
+p^2\frac{\partial}{\partial m_2^2}\frac{\partial}{\partial p^2}F_0^{s_2}=0
 \   \  .
\labbel{p14}\end{eqnarray}

\noindent
From this system one can deduce, eliminating \(m_i^2\) derivatives, that 
\(F_0^{s_2}\) fulfills a very simple differential equation

\begin{eqnarray} 
-p^2{\frac{\partial^2}{\partial (p^2)^2}}F_0^{s_2}
-{\frac{1}{2}}n{\frac{\partial}{\partial p^2}}F_0^{s_2}=0 \ \ .
\labbel{m6} \end{eqnarray} 

Its solution has the following form

\begin{eqnarray} 
F_0^{s_2}={\frac{2S_2(n,m_1^2,m_2^2)}{n-2}}(p^2)^{-{\frac{n-2}{2}}}
 +S_1(n,m_1^2,m_2^2) \ \ ,
\labbel{m7} \end{eqnarray} 

\noindent
where \(S_i(n,m_1^2,m_2^2), \ i=1,2\) are still unknown functions.
Using \Ref{p14} and \Ref{m7} one finds

\begin{eqnarray}
\frac{\partial}{\partial m_1^2}S_1(n,m_1^2,m_2^2)=
\frac{\partial}{\partial m_2^2}S_1(n,m_1^2,m_2^2)=0 \ \ ,
\labbel{p21}\end{eqnarray}

\noindent 
 so the function \(S_1\) does not depend on masses:
 \(S_1(n,m_1^2,m_2^2)=S_1(n)\). This information together with \Ref{p11}
 and \Ref{p12} gives \(S_2(n,m_1^2,m_2^2) = 0\). It means that 
 \(F_0^{s_2}\) is a function of \(n\) only. Its value will be found later on.

Similar, but even simpler, analysis can be done for \(\alpha=s_3\).
From the differential equations with \(m_3\) as an independent variable
one finds that \(F_0^{s_3}\) does not depend on the masses. 
Using that information, from differential equations \Ref{a5} and \Ref{a6} one
finds

\begin{eqnarray} 
p^2{\frac{\partial}{\partial p^2}}F_0^{s_3}=-{\frac{1}{2}}(n-2)F_0^{s_3}
\ \ ,
\labbel{m10} \end{eqnarray} 

\noindent
which gives

\begin{eqnarray} 
F^{s_3}_0=S(n)(p^2)^{-{\frac{n-2}{2}}}\ \ ,
\labbel{m11} \end{eqnarray} 

\noindent
where \(S(n)\) is a function depending only on \(n\).
As the master integrals for \(n=4\) are analytic functions at \(p^2=0\),
the function \(S(n)\) and consequently \(F^{s_3}_0\) have to be identically
equal to zero. As all higher order coefficients  
\(F^{s_3}_i \sim F^{s_3}_0 , \ i=1,\cdots\), the whole series with
\(\alpha=s_3\) vanishes.

The only unknown function \(F_0^{s_2}(n)\) can be found using known
analytical result for \(F_0(n,0,0,m^2,p^2)\) \cite{CCLR1}. Performing
 its expansion for large \(m^2\) and comparing the appropriate terms
 with \Ref{m1},where two masses were set to zero, one finds

\begin{eqnarray} 
  F_0^{s_2}(n)
= {\frac{C^2(n)}{16}}\biggl[ -{\frac{2} {(n-4)^2}} + {\frac{3}{(n-4)}}
 -\left(\frac{7}{2}+\zeta_2\right) + O(n-4)\biggr] \ \ .
\labbel{m13} \end{eqnarray} 

As \(F_0(n,0,0,m^2,p^2)\) was given \cite{CCLR1}
in a form of an expansion around
\(n=4\) the function \(F_0^{s_2}(n)\) is given only in the form of the \(n=4\)
expansion. As this was the last missing part of the expansion one can
find now the complete formula. We report here only leading terms of the
 expansion, but higher order terms can be easily found by algebraic means,
 if necessary. 

\begin{eqnarray}
&&{\kern-10pt}
F_0(n,m_1^2,m_2^2,m_3^2,p^2)= \  \ pole  \ terms  \ in \ (n-4)
\nonumber\\
&&
+C^2(n)\biggr\{
-\frac{1}{16}m_3^2\log^2(m_3^2)
+\frac{3}{16}m_3^2\log(m_3^2)
-\frac{1}{16}m_3^2\biggr(\frac{7}{2}+\zeta(2)\biggr)
\nonumber\\
&&
+\frac{1}{32}\biggr(m_1^2+m_2^2\biggr)\log^2(m_3^2)
\nonumber\\
&&
+\frac{1}{32}\biggr(2m_1^2+2m_2^2+p^2
-2m_1^2\log(m_1^2)-2m_2^2\log(m_2^2)\biggr)\log(m_3^2)
\nonumber\\
&&
+\frac{1}{128}\biggr(-3p^2+16m_1^2\log(m_1^2)+16m_2^2\log(m_2^2)
-4m_1^2\log^2(m_1^2)
\nonumber \\
&&
-4m_2^2\log^2(m_2^2)+(8\zeta(2)-20)(m_1^2+m_2^2)\biggr)
+O\left(\frac{1}{m_3^2},n-4\right)\biggr\}
\labbel{roz1}\end{eqnarray}

The pole terms being identical to the exact pole terms \cite{CCLR1}
are not reported here. As a cross check we have compared the above
 result with the small \(p^2\) expansion of \(F_0(n,m_1^2,m_2^2,m_3^2,p^2)\).
 We have expanded the first two coefficients of the expansion in \(p^2\)
 \cite{CCLR1},
 dependent 
 on \(m_1,m_2,m_3\), assuming
 \(m_3^2 \gg m_1^2,m_2^2\) and found complete agreement between the
 two results. The result
 \Ref{roz1}, which is valid for arbitrary \(p^2,m_1^2\) and \(m_2^2\) provided 
 they are much smaller then \(m_3^2\),
 cannot be however deduced from the small \(p^2\) expansion itself.

\section{Summary}

 A short introduction to the differential equations method used in calculation
 of the scalar multi-loop integrals was presented. A nontrivial large mass
 expansion of the master two-loop sunrise integrals was obtained almost
 completely by algebraic means. The only 'difficult' task, 
 besides solving systems of linear algebraic equations, was to solve
 two simple differential equations \Ref{m6} and \Ref{m10}.
 It shows that the method is extremely powerful and opens new possibilities
 in the field of multi-loop calculations. 
\vskip 1 cm

\noindent
{\bf Acknowledgments}

\vskip 0.3 cm

We would like to thank Michele Caffo for discussion and careful reading of the
 manuscript.

\def\NP{{\sl Nucl. Phys.}} 
\def\PL{{\sl Phys. Lett.}} 
\def\PR{{\sl Phys. Rev.}} 
\def\PRL{{\sl Phys. Rev. Lett.}} 
\def\NC{{\sl Nuovo Cim.}}
\def\AC{{\sl Acta Phys.Pol.}}

\end{document}